# GEOMECHANICAL RISK ANALYSIS FOR THE HISTORICAL SALT CAVERN SAURAU IN "WIELICZKA" MINE

**Witold Pytel**, **Józef Parchanowicz**, **Piotr Mertuszka**,

KGHM CUPRUM Ltd. Research and Development Centre, **Poland**

**ABSTRACT:**

*Based on the finite element method formulated in three dimensions, a selected area of the salt mine has been modeled, embracing totally 2 separated, historically valuable mine rooms. The results of the computer simulations permitted elastic-viscous rock mass stability analyses identifying the areas being more susceptible to damage, presently and in a far future. The geomechanical risk assessment procedure utilized so called safety margins which were defined as a distance between the point characterized by the actual local strain/stress conditions and the instability (limit) surface(s) which location in the 3D stress/strain space could be determined using the well-known strength theories.*

**Key words:** rock mechanics, numerical modelling

## 1. INTRODUCTION

### 1.1. General information

Due to 700-year ore exploitation, within mine field of the royal salt mine "Wieliczka", there has been developed the extremely complex and locally unrecognized structure of chambers and roadways. So far on nine levels of which the lowest is set about 327 m below ground level, there are 2177 chambers localized and marked on maps. The most important central part of a historically developed excavation system, localized on levels I-III between the Kościuszko and the Wilson shafts, includes about 350 historical objects, among them about 40 invaluable monuments of "0" class in different technical conditions.

In the end of 70-ties, when deterioration process caused by significant strata movement within some of the historical objects was observed, an action had been undertaken to preserve them for the next generations. This time, the experts agreed that a backfilling of the most of rooms located below the level V is necessary in order to create stable support for the valuable higher levels of the mine. The backfilling operations involving a huge volume of tailings has been successfully conducted during the years concluding with the rock mass higher stability clearly visible in the regional scale. However, in a smaller scale, most of the individual caverns located at the higher levels of mine, still suffer from time-dependent deformations concluding with spatially confined instabilities. These effects dealt also with the largest caverns, among them the Saurau chamber, are usually due to viscous behavior of surrounding shells of salt rock which are presently controlled using different ground support measures, mostly rock bolts and wooden complex support sets (see Fig. 1). However, for possible future instability potential, appropriate numerical modeling was recommended.

### 1.2. General site description

The Saurau cavern is located within relatively complex mine workings system (Fig. 2), on the IIn (+141 m above sea level), III (+115 m) mine levels with inter-level Lichtenfels (+87 m above sea level). It had been excavated within a one very large ZBt solid salt boulder.

The total cavern's capacity is about 18,000 cubic meters with representative height of about 30 m, length of 50 and width of 20 m. Due to a complex shape, the cavern size depends on the location.

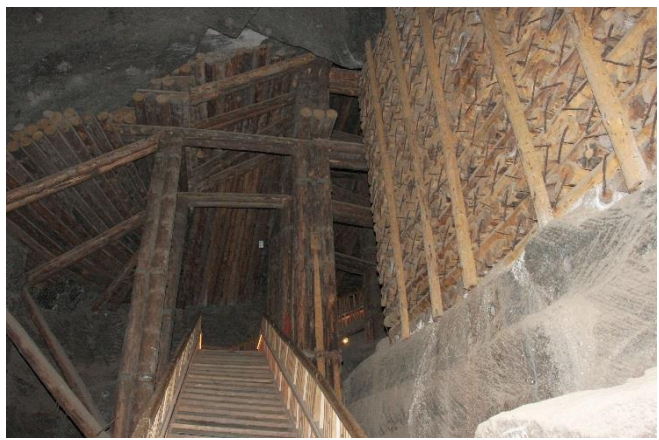

**Figure 1. Support sets' view in Saurau cavern**

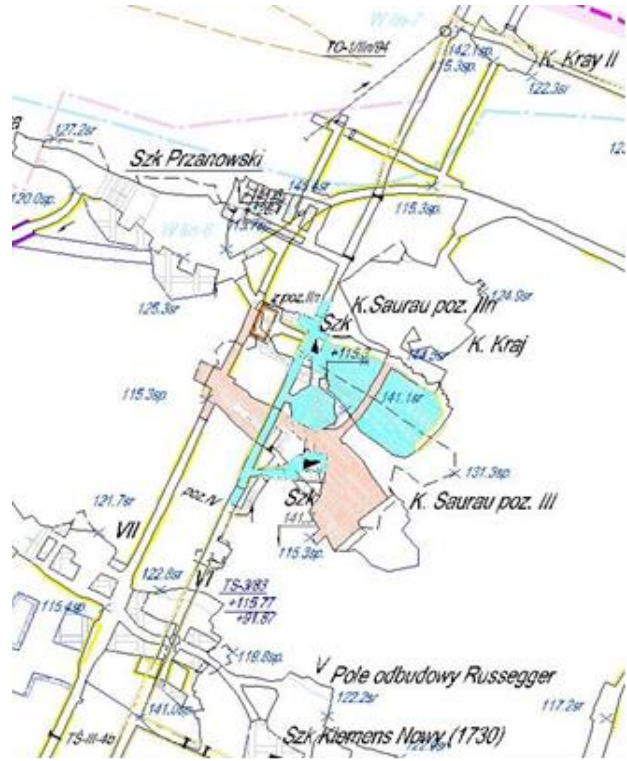

**Figure 2. Saurau cavern at levels II and III**

### 1.3. Site's hydro geological conditions

The Saurau cavern has been excavated in very large green salt rock ZBt lens (Fig. 3-4) associated with large number of smaller salt rock lenses spatially scattered within clayey waste rocks.

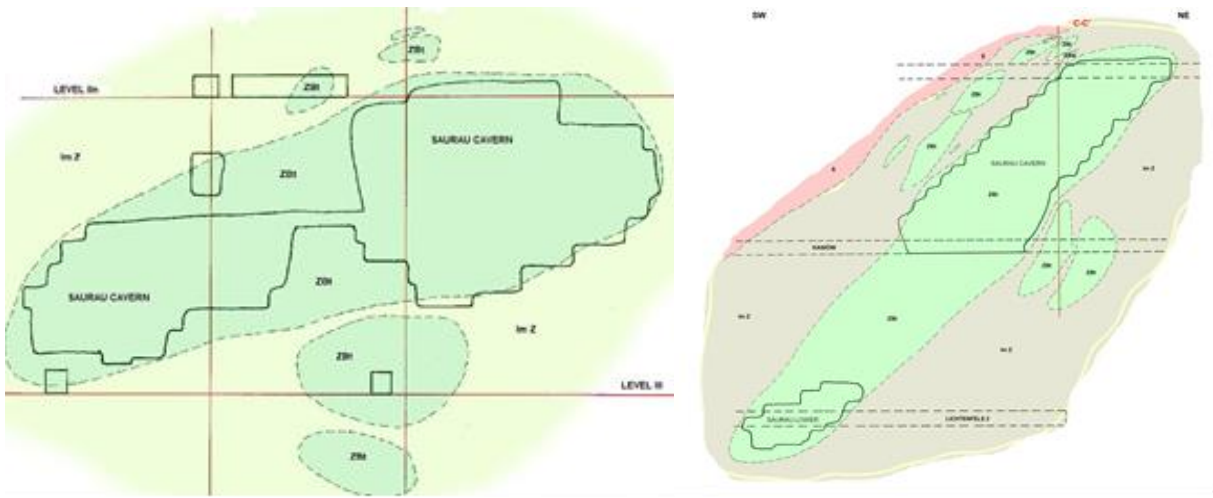

**Figure 3. Vertical cross-sections of Saurau cavern (Przybyło, 2010)**

The analyzed rock mass zone is located under the northern branch of solid salt seam with a dip of about 40 ÷ 45° in NW – SE direction. In this area, at the level IIn, insignificant water inflow is observed, however at the level III and below at Lichtenfels level, no water flow related phenomena has been noticed.

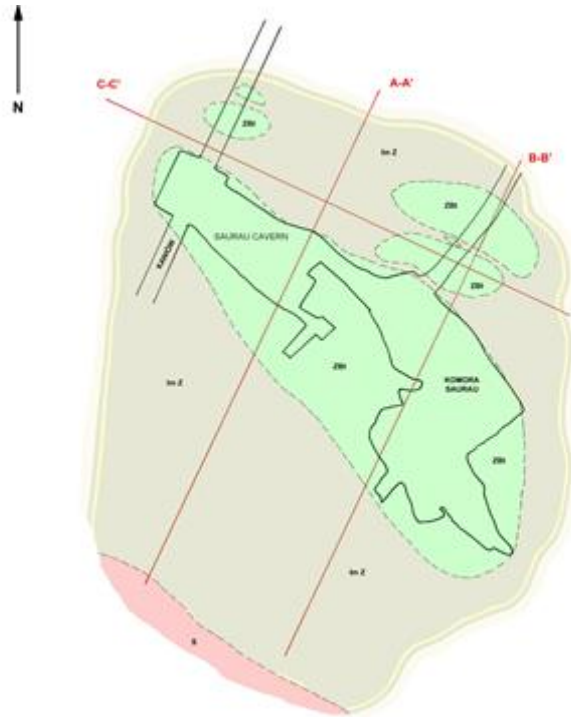

**Figure 4. Saurau cavern horizontal cross-section (Level III) (Przybyło, 2010)**

### 1.4. Current technical state of the Saurau cavern

Taking into account that the analyzed cavern had been excavated in XVII century, its technical state may is considered to be relatively good. Renovation works made during last years were confined to thin shotkrete spraying and epoxy rock bolts installation supplemented by coated steel mesh. Also, wooden and concrete support has been provided in the location where surrounding rock mass has deformed or/and fractured. Today however, due to the cavern's large size and the visitors' safety reasons, additional reinforcement and support has been recommended.

As a part of this activity, numerical modeling of the Saurau cavern and its vicinity has been therefore undertaken with the basic objective to identify the area prone to instability manifested by cracks and visible deformations.

## 2. GEOMECHANICAL ANALYSIS

### 2.1. 3D numerical model of the cavern

Geomechanical problem solution and results visualization were based on the NEi/NASTRAN (2009) computer program code utilizing FEM in three dimensions with the option of creep behavior of selected materials.

In the frame of the presented approach, the different two models have been analyzed:

- The Saurau cavern with including in the close vicinity the presence of the following chambers: Kraj, Przanowski and Długosz,
- The Saurau cavern without including the presence of the Kraj, Przanowski and Długosz chambers, replacing them by large lenses of solid salt rock.

The model (Fig. 5), composed of 285345 finite elements linked by 304900 nodes, was built using available, however rather scarce, historical data confined to several technical and geological drawings.

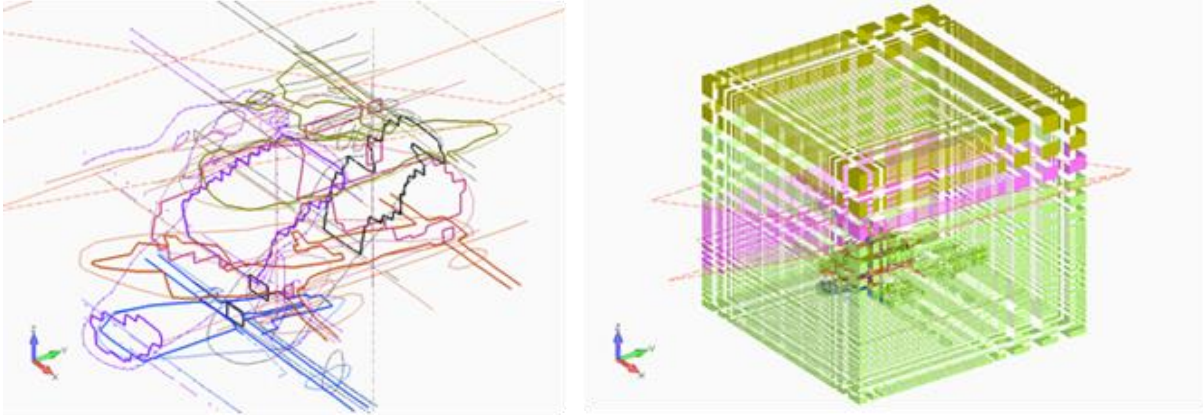

**Figure 5. Drawing skeleton (top) and the numerical FEM model (bottom)**

The model boundary conditions have been formulated as the displacement type limits, while the external load confined to rock mass self-weight was represented by density parameter (gravitation).

Having strength-deformation parameters of the local rock mass (Table 1) as well as results of numerical modeling (stress and deformation fields), one could determine values of so called margins of safety for surrounding rocks based on the Coulomb-Mohr strength theory:

$$F_{cm} = R_c - A \cdot \sigma_1 + \sigma_3 \tag{1}$$

where: $R_c$ – rock mass unconfined compression strength, $\sigma_1$ and $\sigma_3$ – minor and major principal stress, and $A$ – parameter given in Table 1. The negative value of $F_{cm}$ indicates high potential for instability event.

Furthermore, the authors focused on the effective strain and the major principal stress spatial distribution considering them as good instability indicators involving shear and tensile type of mechanisms of failure. The appropriate limit criteria are as follows:

$$\varepsilon_{ef} = \frac{\sqrt{2}}{3}[(\varepsilon_{11} - \varepsilon_{22})^2 + (\varepsilon_{22} - \varepsilon_{33})^3 + (\varepsilon_{33} - \varepsilon_{11})^2 + 6 \cdot (\varepsilon_{12}^2 + \varepsilon_{23}^2 + \varepsilon_{31}^2)]^{0.5} \geq 0.03$$

$$\sigma_1 - R_r \geq 0 \tag{2}$$

where: $R_r$ – tensile strength.

### 2.2. Rock mass geotechnical parameters

From the available database (e.g. Wojnar, 1996) one may conclude that the Saurau cavern is located within relatively soft and weak rock mass composed mostly of marlaceous claystone. In the numerical analysis, three-dimensional geological structure is modeled therefore by introducing three basic material types of different strain-strength and viscous parameters:

- stratiform deposit of outstanding viscous-elastic behavior, composed of stratiform and shaft salt with inclusions of barren rock,
- boulder deposit representing a mixture of marlaceous clays, boulder and granular salt, generally revealing moderate viscous-elastic behavior,
- marlaceous claystone representing all remaining, by assumption purely elastic rocks (claystone, marlaceous claystone, clayey marl, mudstone etc.).

The average strength-deformation parameters of rock mass located nearby of considered cavern and afterwards used in the numerical analyses, have been presented in Table 1.

Table 1. Average values of geotechnical parameters of rocks

| Kind of Rock | $R_c$ (MPa) | $R_r$ (MPa) | Modulus of deformation E (MPa) | Poisson's ratio ν | Angle of internal friction φ | A | Coefficient of viscosity η (Pa sec) |
|---|---|---|---|---|---|---|---|
| Tertiary and Quaternary deposits | | | 70 | | | | |
| Stratiform and shaft salt | 40 | 3.5 | 180 | 0.4 | 40 | 4.6 | 4.2E17 |
| Boulder salt deposits | 37 | 2.4 | 150 | 0.4 | 37 | 4.02 | 4.2E17 |
| Marlaceous claystone + 10% shaft salt | 5 | 0.3 | 60 | 0.3 | 30 | 3.0 | 4.2E18 |

Creep properties of salt rock, particularly of stratiform and shaft salt, are considered to be the most important factor inducing the presently observed rocks' movement within the Saurau cavern. This uniformly increasing with time movement reminds typical behavior of the Maxwell liquid model. The creep parameter η of a such simple model has been determined based on field and laboratory measurements and tests.

### 2.3. Geomechanical analyses of the Saurau cavern and its vicinity

*2.3.1. Model calibration*

The model calibration has been done assuming that roof-floor convergence developed for the last year within the cavern is equal to actually measured of 2 mm distance. In this case, the value of creep parameter of marlaceous claystone has been established as the decision parameter. This implied that the marlaceous claystone contained the of salt additive of a given amount. The successive phases of modeling and computation process with gradually increasing salt ratio, assumed the Newton liquid creep parameter decrement according to the function:

$$\eta = \frac{4{,}5 \cdot 10^{17}}{Z_z} \quad (3)$$

where: $Z_z$ – volumetric amount of salt within the claystone, $4.5E17$ – creep parameter of salt deposits accepted as an average value from the appropriate literature (Slizowski, 2006; Wojnar et al., 1996). Since the assessed value of roof-floor convergence reached 2 mm/year for $Z_z$ = 0.1 (10%), in basic geomechanical computations, creep coefficient for marlaceous claystone $\eta$ = 4.5 E18 Pa • sec has been set up.

*2.3.2. Results of numerical computations*

The numerical model of the Saurau and Kraj caverns shown in Figure 6 consisted of 37 layers along which the obtained results have been presented in the following categories of variables: yearly x-y-z translations increment, stresses in x-y-z coordinate system, values of the effective strain, and safety margin Fcm in 3D space.

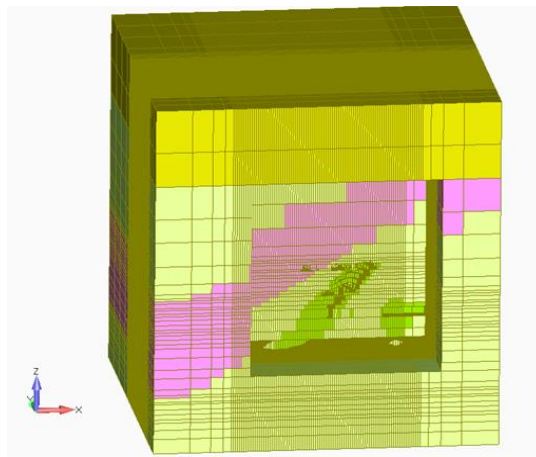

**Figure 6. General view of the numerical model**

*2.3.3. Yearly translation increment*

Calculation results have confirmed that the yearly increment of vertical displacement $w_z$ (see Fig. 7) will not be greater than 0.003 m and are in agreement with values observed in the nature, accepted as the base for the model calibration process. Calculated values of horizontal deformation increment $w_x$ i $w_y$ may reach no more than ± 0.0005 m.

One may notice that rock mass located at southern part of the cavern tends moving in northern direction while increment of displacements in E-W direction is negligible.

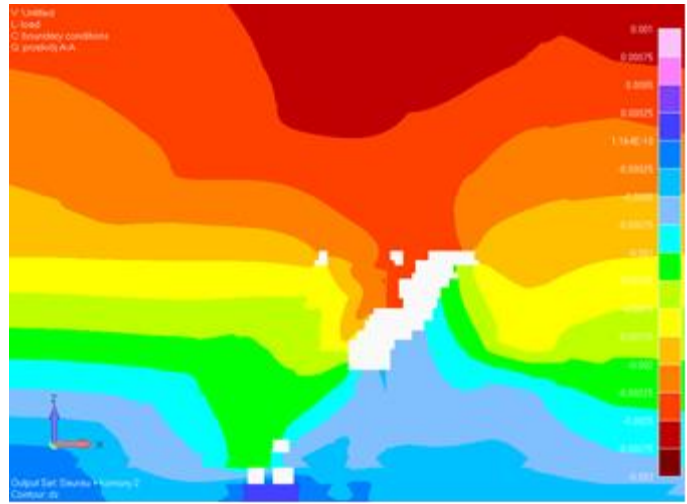

**Figure 7. Yearly increment (m) of vertical displacement contour (cross-section A-A)**

Although the calculated value of vertical translation at the model's surface reaches value of 0.00275 m, one may assume that his number is underestimated due to the limited volume of the analyzed model.

*2.3.4. Stresses in x-y-z coordinate system*

The obtained calculation results have revealed that horizontal normal stresses $\sigma_x$ i $\sigma_y$ (Fig. 8) reach relatively low, negative values however, one may identify very limited areas of positive stresses located at the opened cavern's surface. This may suggest tendencies for scaling development at limited range.

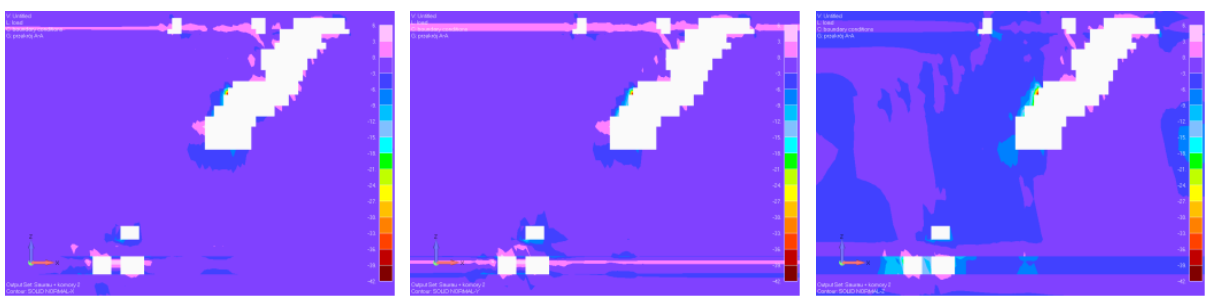

**Figure 8. Normal stresses (MPa) contours at A-A cross-section ($\sigma_x$ – left, $\sigma_y$ – middle, $\sigma_z$ – right)**

Also vertical stress $\sigma_z$ (see Fig. 8 right) reaches moderate values of about 2 – 4 MPa however the areas of positive stress may be noticed at the cavern's surface (scaling). Also, southern wall of the cavern located at the depth of 123 m is locally compressed due to the close presence of isolated salt lenses. This part of the cavern is reinforced already by rock bolts.

Calculated shear stresses contours (Fig. 9) indicate that $\sigma_{xy}$ stress reaches generally values of very low magnitude and therefore they may be considered in geomechanical analyses to be a parameter of second order importance. Contours of remaining categories of shear stresses indicate however that the cavern's southern corner, at a depth of 123 m, is subjected to significant shearing and therefore in this area (also in the lower part of the cavern) one may expect fractures development.

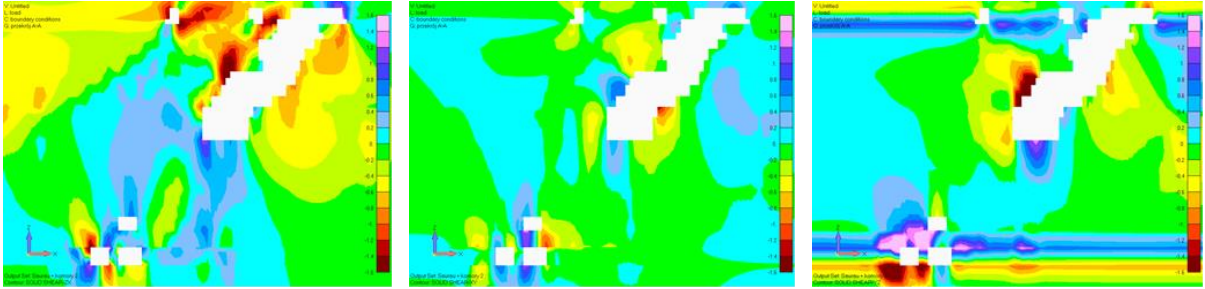

**Figure 9. Shear stresses (MPa) contours at cross-section A-A ($\sigma_{xz}$ – left, $\sigma_{xy}$ – middle, $\sigma_{yz}$ – right)**

*2.3.5. Principal stress $\sigma_1$*

The calculated minor principal stress $\sigma_1$ when reveals positive value may be considered as a good indicator of tendency to rock mass instability occurrence realized mostly by a tensile type mechanism of failure – open fractures.

Analyzing Figures 10-11 from this point of view, one may find that the areas of a positive $\sigma_1$ value are confined to the cavern walls of selected location (potential for scaling) or to a kind of the lower caverns projection formed in a characteristic pattern following these caverns' shape (see lower Saurau). Therefore open fractures development may be expected there.

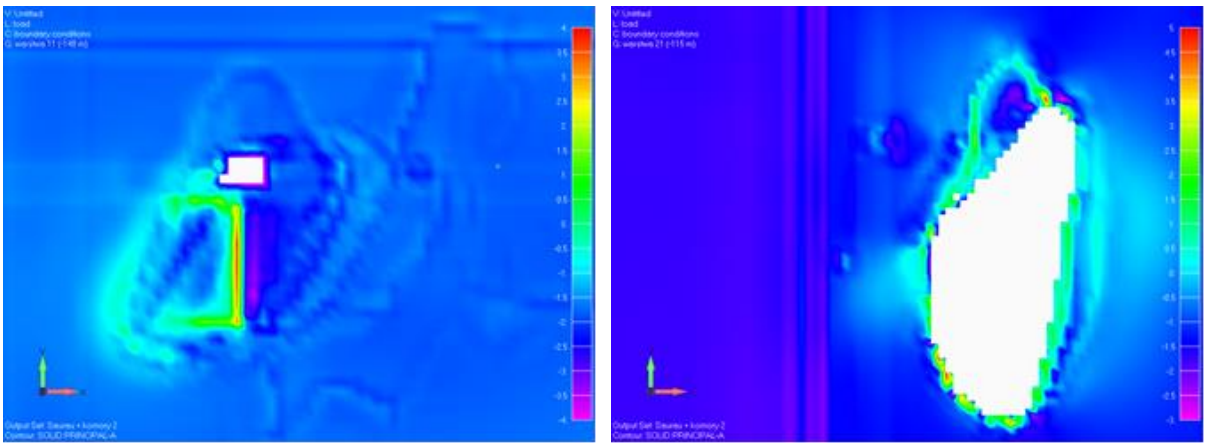

**Figure 10. Principal stress $\sigma_1$ contours (MPa) at 148 m (left) and 115 m below surface (right)**

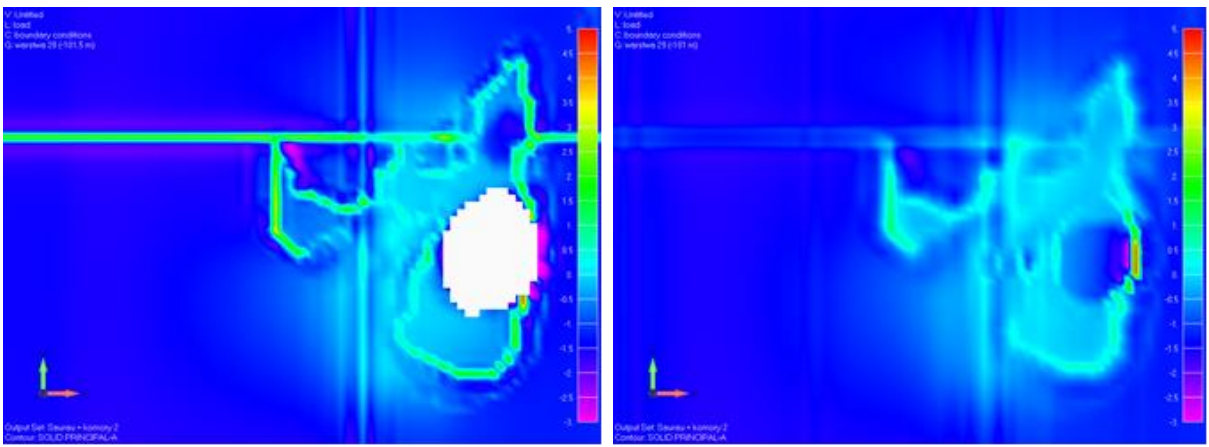

**Figure 11. Principal stress $\sigma_1$ contours (MPa) at 101.5 m (left) and 101.0 m below surface (right)**

### 2.3.6. Effective strain $\varepsilon_{ef}$

The effective strain (see Eq. 2), when is of higher value, may serve as a kind of indicator of the areas where a high potential for rock mass separation and its fracturing exists. When analyzing Figures 12-13 from this point of view, one may conclude that rock mass in the Saurau cavern area is relatively intensively loaded, particularly along surfaces of the contact between marlaceous claystone and boulder salt deposits. Therefore one may expect that in those locations, the boulder salt elements shall work partially independently of surrounding rocks, forming a kind of blocked, distinct salt elements structure of lower potential for stability than that characteristic for the intact rock mass.

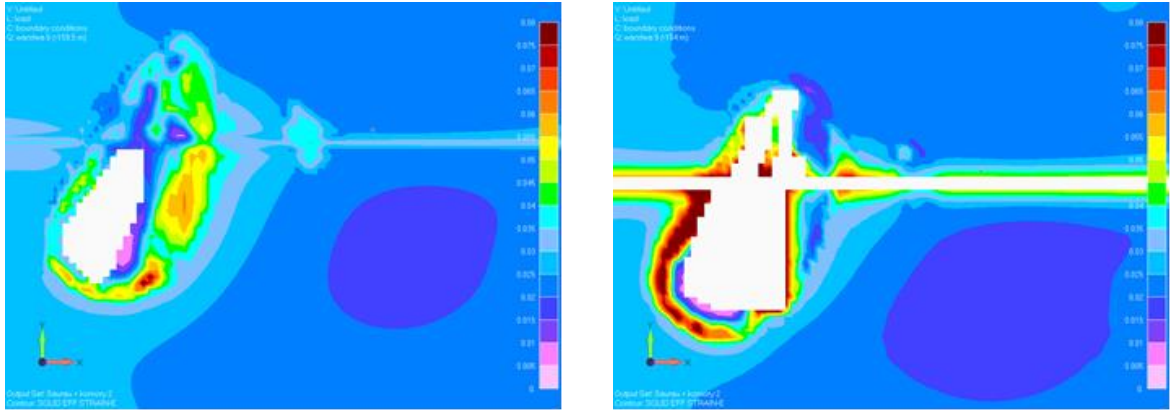

**Figure 12. Effective strain contours at 159.5 m (left) and 154 m (right) below the ground surface**

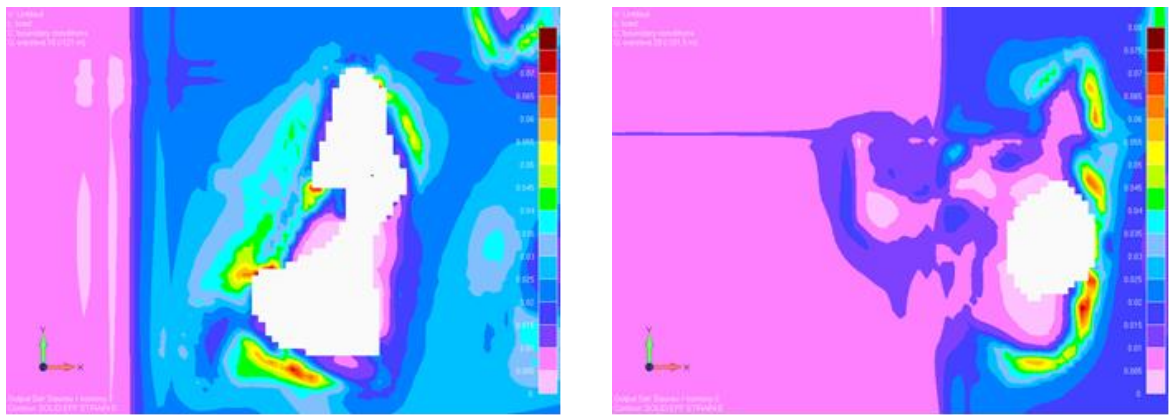

**Figure 13. Effective strain $\varepsilon_{ef}$ contours at 121.0 m (left) and 101.5 m (right) below surface**

The calculated yearly increment of effective strain (see Fig. 14) confirms insignificant changes comparing with the initial state.

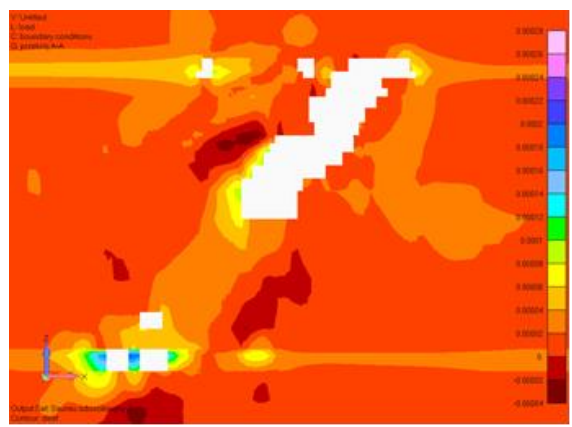

**Figure 14. Effective strain increment contour (cross-section A-A)**

*2.3.7. Safety margin $F_{cm}$*

Negative values of calculated safety margin $F_{cm}$ (Fig. 15) indicate higher hazard from instabilities caused by shear stress overloading (Coulomb-Mohr strength hypothesis). The areas of transition between marlaceous claystone and salt deposits seem to be the most sensitive from this point of view however, since safety margin values are generally greater than zero, this kind of risk may be presently treated rather as an irrelevant hazard.

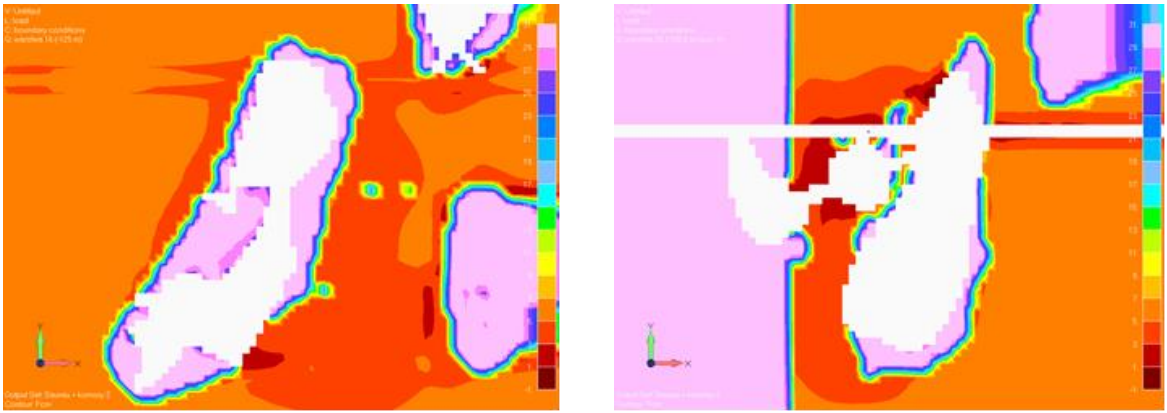

**Figure 15. Safety margin $F_{cm}$ (MPa) contours at 125.0 m (top) and 105.5 m (bottom) below surface**

### 3. CONCLUSIONS

The comparative analytical material obtained using numerical modeling of the Saurau cavern and its vicinity for the present moment and for the future (365 days ahead), clearly indicate that the first importance causes of observed rock mass deformations are:

- viscous behavior of the most rocks composing the geological environment of the „Wieliczka" mine, and
- rock mass structure within the immediate vicinity of the caverns, where one may observe the area of transition between weak marlaceous claystone and stronger boulder salt structures weakened however by excavated caverns' presence.

Therefore in an extreme case one may encounter a shell-type structure which may break in bended parts and corners and afterwards works as a structure composed of distinct elements. Due to the high shear stress values, these distinct elements may develop mostly at the contacts of materials revealing strain properties of significant differences. Presented geomechanical analyses hale proved that instability hazards decrease very rapidly with the distance from the cavern's wall. This permits effective applying of rock bolts type, shotcrete and other means of ground support systems. The sensitive points of Saurau cavern which were identified above using numerical approach, may be then treated with a such stability tools (see Fig. 16).

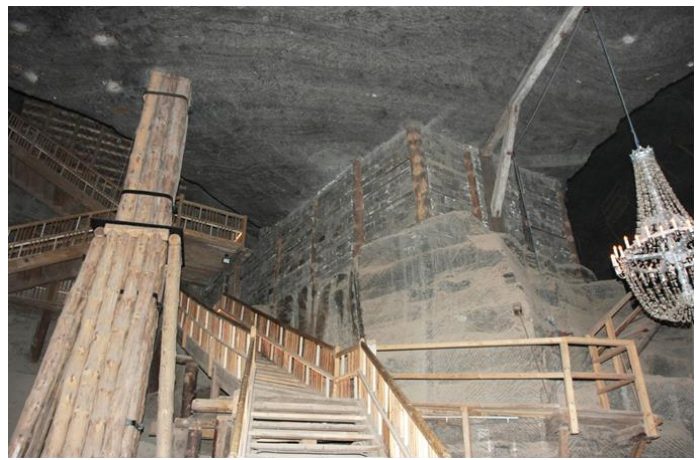

**Figure 16. Supported salt roof strata in Saurau cavern**